\documentclass[app]{revtex4-1}

\usepackage{graphicx}
\usepackage{multirow}
\usepackage{makecell}
\usepackage{mathtools}
\usepackage{amsmath}
\usepackage{siunitx}
\usepackage{upgreek}
\usepackage{epstopdf}
\draft 

\begin{document}


\title{Ultra-high Q terahertz whispering-gallery modes in a silicon resonator} 



\author{Dominik Walter Vogt}
\email[Author to whom correspondence should be addressed.  Electronic mail: ]{d.vogt@auckland.ac.nz}
\affiliation{Dodd-Walls Centre for Photonic and Quantum Technologies, Department of Physics, The University of Auckland, Private Bag 92019, Auckland 1142, New Zealand}
\author{Rainer Leonhardt}
\affiliation{Dodd-Walls Centre for Photonic and Quantum Technologies, Department of Physics, The University of Auckland, Private Bag 92019, Auckland 1142, New Zealand}

\date{\today}

\begin{abstract}
We report on the first experimental demonstration of terahertz (THz) whispering-gallery modes (WGMs) with an ultra high quality (Q) factor of $1.5 \times {10}^{4}$ at 0.62\,THz. The WGMs are observed in a high resistivity float zone silicon (HRFZ-Si) spherical resonator coupled to a sub-wavelength silica waveguide. A detailed analysis of the coherent continuous wave (CW) THz spectroscopy measurements combined with a numerical model based on Mie-Debye-Aden-Kerker (MDAK) theory allows to unambiguously identify the observed higher order radial THz WGMs.    
\end{abstract}


\maketitle 

\section{Introduction}
WGM resonators are well established in the optical regime and provide exciting opportunities in fundamental research and technical applications alike \cite{matsko2006optical,LPOR:LPOR200910016,Foreman:15}. The small dimensions and the achievable extremely high Q factors render WGM resonators appealing for many applications, e.g. spectral switches and filters, optical delay lines, lasers and sensors to name but a few \cite{ilchenko2006optical,yang2014quasi}.

However, up to now WGM resonators occupy only a niche in the THz frequency range, albeit their interesting characteristics. In particular, the long wavelength in the THz frequency range with hundreds of microns significantly eases the requirement for ultra high precision machining of the resonator structure present at optical frequencies. However, due to the usually strong material absorption present at THz frequencies, the choice of the resonator material is essential to achieve high Q THz WGMs. Also, most low-loss dielectric materials tend to have an increasing material absorption with frequency in the THz frequency range, hampering the breakthrough of THz WGMs at higher frequencies, i.e. above 0.3\,THz \cite{naftaly2007terahertz}. High density Polyethylene (HDPE) is one of the promising materials for THz WGM resonators. HDPE disc resonators with a diameter of 30\,mm and 10.5\,mm have been reported to show high loaded Q factors of 3000 at 0.24\,THz \cite{annino1997whispering} and 1000 at 0.3\,THz \cite{
Preu13}, respectively. Also, a loaded Q factor of 800 at about 0.25\,THz has been achieved in a quartz disc resonator with a diameter of 3.62\,mm \cite{Preu:08}. We recently demonstrated a THz WGM bubble resonator made of quartz glass with a Q factor of 440 at 0.47\,THz for critical coupling. The bubble resonator provides a particularly interesting platform for sensing due to the closed environment of the bubble design \cite{Vogt:172}. Finally, another obvious choice for the realization of high Q THz WGM resonators is high resistivity float zone silicon (HRFZ-Si). While machining and coupling into such a high refractive index resonator material poses its own issues (as discussed below), HRFZ-Si provides the striking advantage of a very low material absorption over a wide frequency range. Moreover, silicon is an ideal material for the realization of integrated THz WGM resonator systems. A cylindrical resonator made of HRFZ-Si with a diameter of 5\,mm was first used to observed the propagation of sub-
picosecond 
WGM THz pulses \cite{Zhang:03}. A spherical HRFZ-Si WGM resonator was first reported by our group, with a Q factor of 1600 at 0.35\,THz for critical coupling \cite{Vogt:173}.

In this work, we report on the first experimental observation of ultra high Q THz WGMs with a Q factor of $1.5 \times {10}^{4}$ for critical coupling at 0.62\,THz in a 4\,mm diameter spherical HRFZ-Si resonator. To the best of our knowledge, the achieved Q factor of $1.5 \times {10}^{4}$ exceeds by far any other reported resonant structure in the THz frequency range \cite{nagel2006thz,cao2012low,yee2009high,Li:17,mendis2009comparison,bingham2008terahertz,mendis2009terahertz,george2007integrated,xu2016high,Yang:17,PhysRevLett.112.183903}. Amongst the resonant structures with the highest Q factors reported in literature are, e.g., a bragg grating in a parallel plate waveguide (Q=\,436) \cite{nagel2006thz}, a silicon photonic crystals slab (Q=\,1020) \cite{yee2009high}, a plasmonic Fano metamaterial (Q=\,227) \cite{cao2012low}, a 3D-printed hollow core bragg waveguide with defect layer (Q=\,55) \cite{Li:17}, and a HDPE disc WGM resonator (Q=\,3000) \cite{annino1997whispering}. The ultra high Q factor 
achieved with the Si sphere is a significant step towards the realization of THz photonics technologies based on WGM resonators as they are already well established in the optical regime. In particular, the implementation of a highly sensitive integrated WGM Si sensor at THz frequencies is within reach.

Unexpectedly, critical coupling of the WGMs is easily achieved via evanescent coupling from a sub-wavelength waveguide made of silica glass. HRFZ-Si (n=3.42) and silica glass (n=1.96) demonstrate a significant discrepancy in the material refractive indices, and therefore seemingly prevent the phase matching criteria for evanescent coupling \cite{matsko2006optical}. We demonstrate that the phase refractive indices can be matched close to the surface of the HRFZ-Si resonator, and therefore critical coupling to the THz WGMs can be achieved.

\section{Methods}

\begin{figure}[b]
\centering
\includegraphics[width=13.5cm]{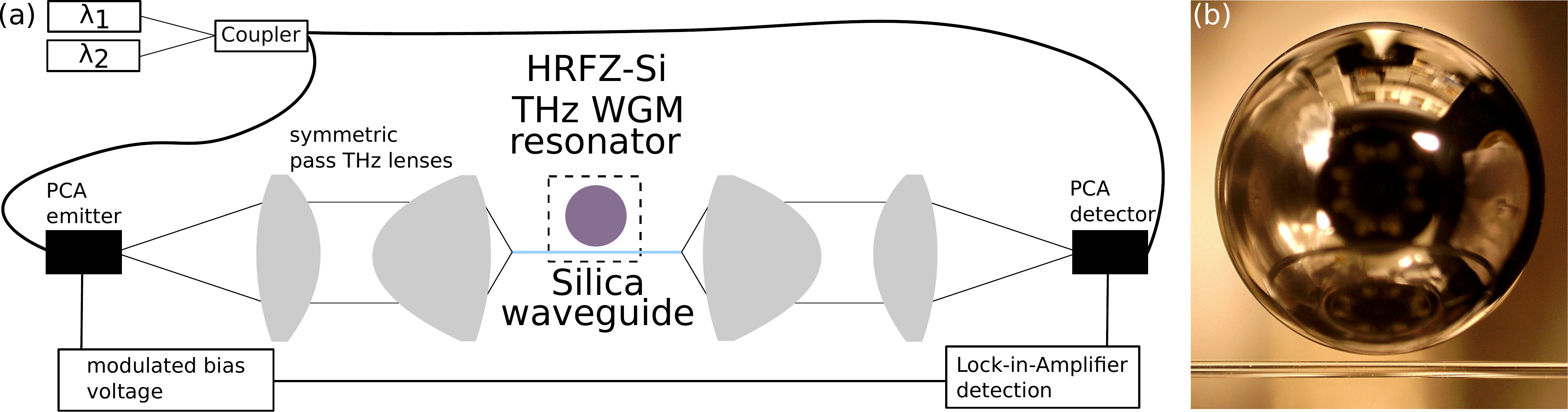}%
\caption{(a) schematic of the CW THz spectroscopy setup using symmetric pass polymer lenses to couple into the silica waveguide (in light blue), and (b) microscope image of the 4\,mm diameter HRFZ-Si sphere (top) and 200\,$\upmu$m diameter silica waveguide (bottom) taken in the black dashed box indicated in (a).\label{Fig0}}%
\end{figure}

The material absorption of the resonator is by far the most dominant factor limiting the Q factor of THz WGMs. HRFZ-Si (resistivity $>\,10\,\mathrm{k\Omega cm}$) has a very low material absorption ($<\,0.025\,{\mathrm{cm}}^{-1}$) and a nearly constant refractive index of 3.416 in the frequency range from 0.4\,THz to 2\,THz \cite{Dai:04}. Therefore, HRFZ-Si is perfectly suited for the implementation of a high Q THz WGM resonator. Also, the basically frequency independent refractive index of HRFZ-Si facilitates an easier identification of the experimentally observed WGMs as discussed in detail below. To numerically analyze the resonance frequencies of the HRFZ-Si resonator, the diameter of the sphere has to be known with great precision. As a micrometer with 1\,$\upmu$m precision has not shown any deviation from the spherical shape, the diameter of the resonator has been determined by weighing the sphere with a 0.1\,mg precision scale. The calculated diameter of the sphere is 3.9974\,$\pm$\,\num{4e-4}\,mm (
density of silicon: 2.329037\,$\pm$\,\num{1e-6}\,g/${\mathrm{cm}}^{3}$ \cite{waseda2004density}). In contrast to optical WGM resonators, imperfections in the resonator shape or surface roughness are usually negligible due to the large wavelength in the THz frequency range compared to the precision of the fabrication processes, especially for a spherical shape. 

The waveguide is a 200\,$\upmu$m diameter air-silica step index fiber made of low OH content silica glass. The waveguide was obtained by removing the cladding of a commercially available multi-mode optical fiber. The low OH content silica glass has a relative low material absorption ($\approx$\,0.2\,${\mathrm{cm}}^{-1}$), and a nearly constant refractive index of 1.96 in the frequency range from 0.3\,THz to 1\,THz \cite{material}. Due to the sub-wavelength cross-section of the waveguide higher order modes are not supported (the calculated V parameter at 635 GHz is 2.2 \cite{agrawal2007nonlinear}), and the fundamental mode has a substantial evanescent field. The latter reduces the losses due to material absorption, and is particularly important for the coupling between the waveguide mode and the WGMs of the resonator.

Experimentally, the WGMs of the HRFZ-Si resonator are characterized using coherent CW THz spectroscopy based on heterodyne detection (Toptica TeraScan 1550nm \cite{Deninger2015}). A schematic of the experimental setup can be seen in Fig. \ref{Fig0}. The THz radiation is coupled into the sub-wavelength waveguide with specially developed symmetric-pass polymer lenses with a very small focal length of 25\,mm and a numerical aperture (NA) of 1 \cite{lens}. A focal spot size in the order of 0.6\,${\lambda}_{0}$ facilitates efficient coupling to the silica waveguide (NA$\approx$1.69). The polarization of the THz electric field is parallel to the resonator surface (out of the plane in Fig. \ref{Fig0}), therefore only transversal  electric (TE) WGMs are excited.

To spectrally characterize the WGMs, the THz radiation transmitted through the waveguide coupled to the WGM resonator is detected, and normalized to the waveguide transmission without the resonator in close vicinity. The position of the resonator relative to the waveguide is controlled via a 3D computer-controlled translation stage with a precision of 0.2\,$\upmu$m. The HRFZ-Si sphere is mounted on a thin aluminum post to avoid any distortion in the equatorial plane of the WGM. The measured phase modulated photo current is directly proportional to the transmitted THz radiation, and is further analyzed using Hilbert transformation. The analysis based on Hilbert transformation significantly improves the effective frequency resolution achievable with the CW THz spectroscopy measurements. A detailed description of the data analysis as well as a comparison to the traditional analysis is provided in our previous publication \cite{Vogt:17}. The amplitude and phase information of the coherent detection 
scheme provide detailed information about the spectral 
properties of the analyzed WGMs. In particular the phase information allows to unambiguously determine the coupling state of the WGMs, while the amplitude information allows to determine the Q factor of the WGMs. 

The measured resonance frequencies and loaded Q factors at critical coupling of the THz WGMs are compared to numerical results based on Mie-Debye-Aden-Kerker (MDAK) theory \cite{Hightower:88}. Knowing the frequency dependent complex refractive index of the HRFZ-Si resonator material and the resonator radius, the resonance frequencies as well as the Q factors of the WGMs can be derived from the spectral information of the partial wave coefficients computed with MDAK theory according to \cite{Hightower:88}. Alternatively, if the radius of the resonator under study and the experimentally determined free spectral ranges (FSRs) of the WGMs are well known, the MDAK theory can be used to determine the complex refractive index as a function of frequency. Please note, that the provided Q factors obtained from MDAK theory are unloaded Q factors, which include radiation losses (similar to bend losses for waveguides) and losses due to material absorption, but not contributions due to the evanescent coupling. As the 
unloaded Q factors are a factor of two larger than the ones obtained at critical coupling \cite{Gorodetsky:99}, the material loss for HRFZ-Si (see discussion below) in the MDAK simulations was adjusted so that the simulations gave twice the value of the experimentally obtained loaded Q factors at critical coupling.

Finally, to study the phase matching criteria for the evanescent coupling between the HRFZ-Si resonator and the silica waveguide, we numerically determine the phase refractive indices of the fundamental mode guided in the sub-wavelength waveguide. The phase refractive indices are obtained from the solution of the eigenvalue equation for a standard step-index waveguide (air-silica) \cite{agrawal2007nonlinear}. 



\section{Mode identification}

\begin{figure}[b]
\centering
\includegraphics[width=10cm]{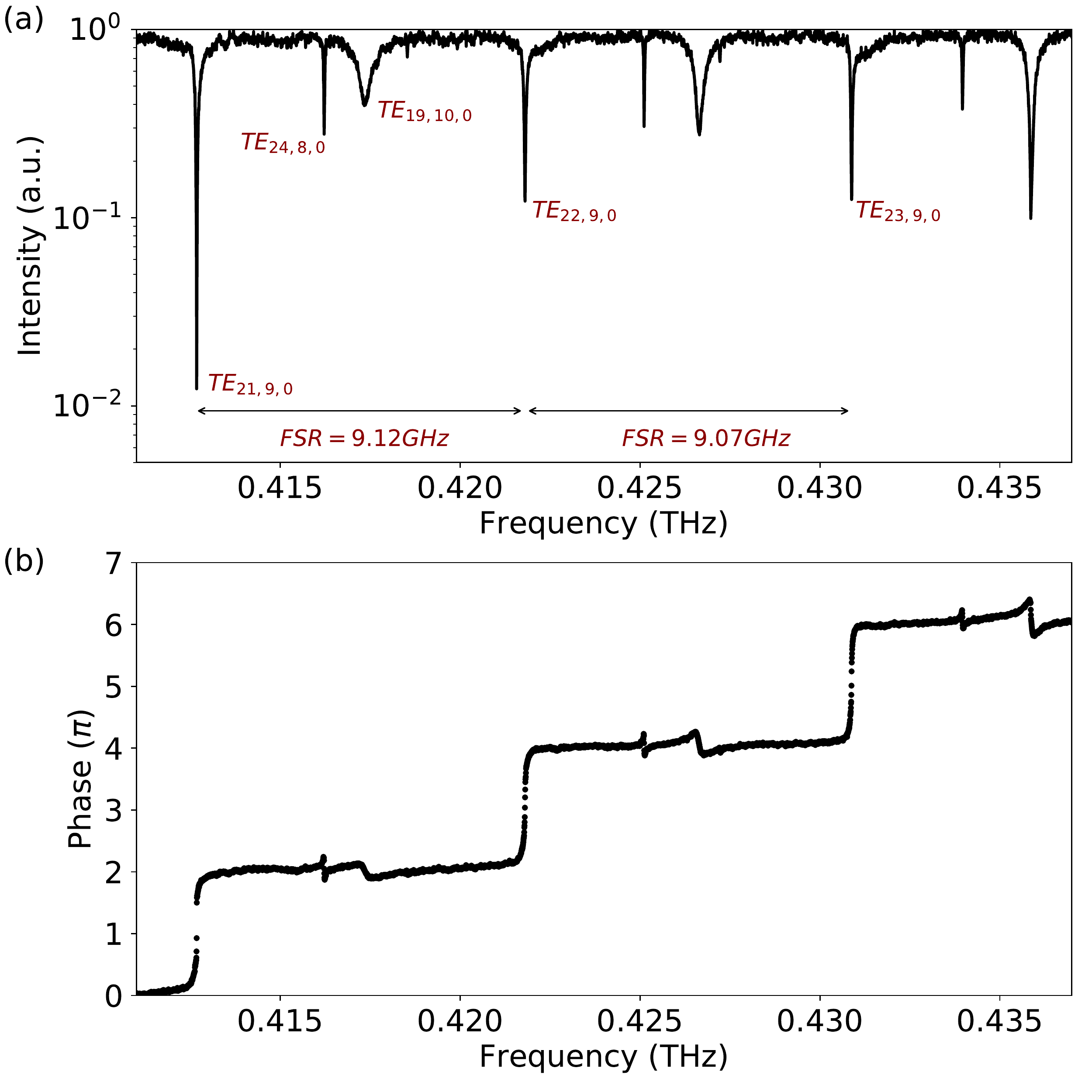}%
\caption{(a) normalized measured intensity transmission of the HRFZ-Si resonator coupled to the single mode silica waveguide in the frequency range from 411\,GHz to 437\,GHz on a logarithmic scale. The first mode of each shown mode family is labeled in the graph. The FSR of the 9th higher order radial mode family is indicated with the horizontal arrows, and (b) corresponding phase profile of the WGMs shown in (a).\label{Fig1}}%
\end{figure}

\begin{figure}[t]
\centering
\includegraphics[width=10cm]{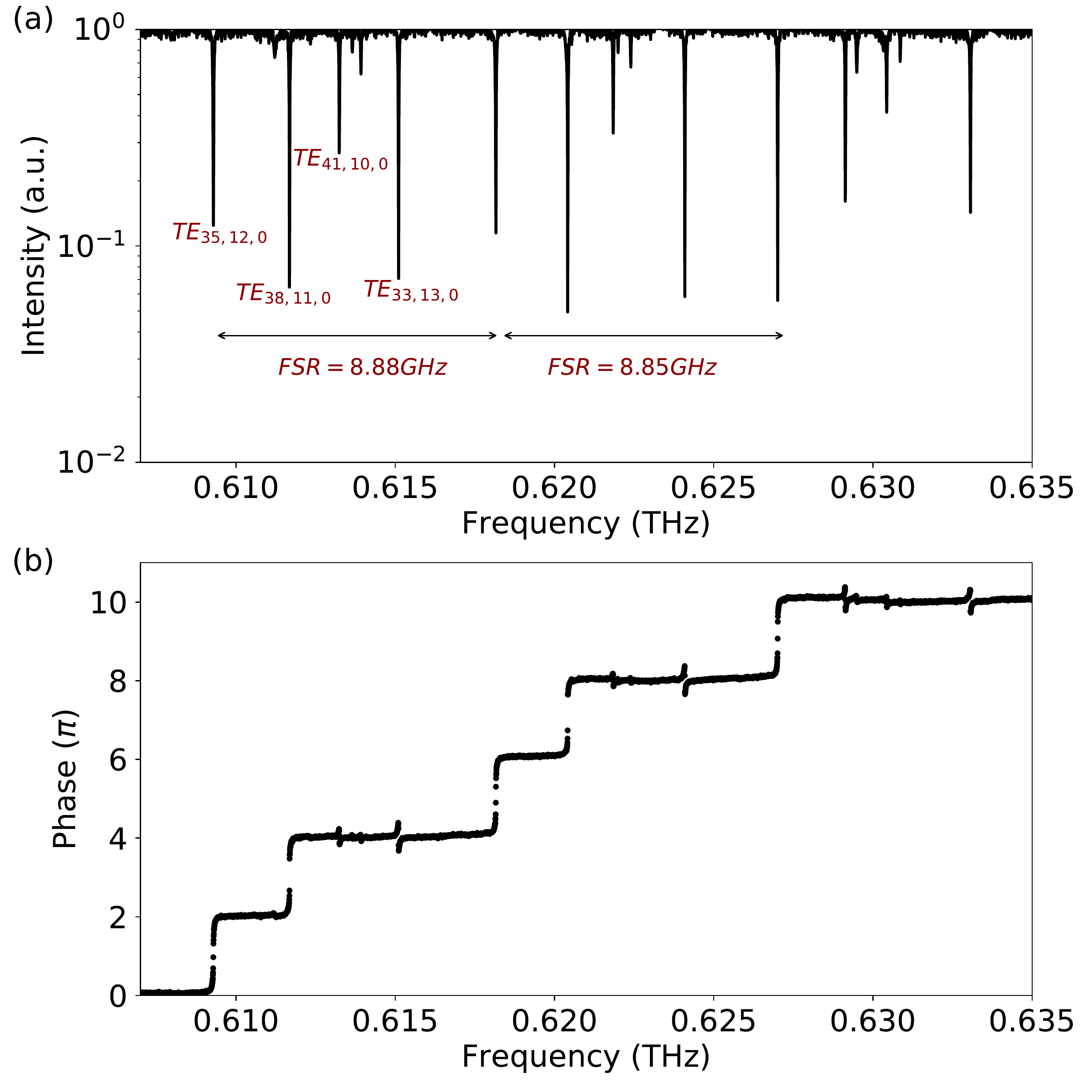}%
\caption{(a) normalized measured intensity transmission of the HRFZ-Si resonator coupled to the single mode silica waveguide in the frequency range from 607\,GHz to 635\,GHz on a logarithmic scale. The first mode of each shown mode family is labeled in the graph. The FSR of the 12th higher order radial mode family is indicated with the horizontal arrows, and (b) corresponding phase profile of the WGMs shown in (a). \label{Fig2}}%
\end{figure}

Fig. \ref{Fig1} and Fig. \ref{Fig2} show the normalized intensity transmission on a logarithmic scale (a) and the phase profile (b) of the silica waveguide coupled to the HRFZ-Si resonator measured in the frequency ranges from 411\,GHz to 437\,GHz and 607\,GHz to 635\,GHz, respectively. The frequency step size in the scans is 4\,MHz. Both frequency ranges clearly demonstrate the presence of WGM resonances, but also show distinct features in each frequency range. An essential step in explaining the spectra is to unambiguously identify the experimentally observed WGMs. The identification of the modes does not only improve the overall understanding of the system, but is also a crucial step for many metrology applications like sensing or material characterization.

The WGMs can be identified from the spectra by comparing the absolute frequencies and the FSRs of corresponding mode families with the numerically predicted values. Hereby a mode family refers to WGMs with the same number of nodes in the radial direction of the resonator (the nomenclature used throughout this work is described in detail below). However, the absolute frequencies are more prone to errors compared to the FSRs, e.g. due to fluctuations of the absolute frequency in the CW THz spectroscopy measurements or because of a small red shift introduced by the waveguide (as discussed below). The FSR is unique for spectrally neighboring WGMs in one mode family, and any shifts in the absolute frequencies are canceled out. Therefore the following discussion about the mode identification will concentrate on the FSRs of the HRFZ-Si resonator rather than the absolute frequencies. Nonetheless, as discussed below, both the absolute frequencies and the FSRs from the measurements show excellent agreement with the 
numerical theory.         

\begin{table}[h]
\caption{Comparison of the experimentally obtained resonance frequencies (${\mathrm{f}}_{\mathrm{exp}}$) and FSRs (${\mathrm{FSR}}_{\mathrm{exp}}$) with the numerical results from the MDAK theory (${\mathrm{f}}_{\mathrm{MDAK}}$, ${\mathrm{FSR}}_\mathrm{{MDAK}}$) in the frequency range from 411\,GHz to 437\,GHz. The rows are clustered according to their radial order, and the corresponding WGMs are labeled in the second column. The last column shows the calculated effective refractive index of the WGMs at the surface of the resonator (${\mathrm{n}}_{\mathrm{ph}}$\raisebox{-.2em}{$\vert_{{r}_{0}}$}).\label{table1} }
\begin{center}
\begin{tabular}{ |c|c|c|c|c|c|c| } 
\hline
\makecell{radial order\\ $q$}& Mode & ${\mathrm{f}}_{\mathrm{exp}} [GHZ]$ & ${\mathrm{f}}_{\mathrm{MDAK}} [GHZ]$  & \makecell{${\mathrm{FSR}}_{\mathrm{exp}}$\\ $[GHZ]$} & \makecell{${\mathrm{FSR}}_{\mathrm{MDAK}}$\\ $[GHZ]$} & ${\mathrm{n}}_{\mathrm{ph}}$\raisebox{-.2em}{$\big\vert_{{r}_{0}}$} \\
\hline
\multirow{3}{4em}{\centering 8} & ${\mathrm{TE}}_{24,8,0}$ & 416.25 & 416.37 & \multirow{2}{4em}{\centering 8.89} & \multirow{2}{4em}{\centering 8.88} & 1.376\\ 
& ${\mathrm{TE}}_{25,8,0}$ & 425.14 & 425.25 & \multirow{2}{4em}{\centering 8.85} & \multirow{2}{4em}{\centering 8.85} & 1.403\\ 
& ${\mathrm{TE}}_{26,8,0}$ & 433.99 & 434.10 &  &  & 1.430\\ 
\hline
\multirow{3}{4em}{\centering 9} & ${\mathrm{TE}}_{21,9,0}$ & 412.71 & 412.82 & \multirow{2}{4em}{\centering 9.12} & \multirow{2}{4em}{\centering 9.12} & 1.214\\ 
& ${\mathrm{TE}}_{22,9,0}$ & 421.83 & 421.94 & \multirow{2}{4em}{\centering 9.07} & \multirow{2}{4em}{\centering 9.07} & 1.245\\ 
& ${\mathrm{TE}}_{23,9,0}$ & 430.90 & 431.01 &  &  & 1.274\\ 
\hline
\multirow{3}{4em}{\centering 10} & ${\mathrm{TE}}_{19,10,0}$ & 417.34 & 417.46 & \multirow{2}{4em}{\centering 9.29} & \multirow{2}{4em}{\centering 9.27} & 1.087\\ 
& ${\mathrm{TE}}_{20,10,0}$ & 426.64 & 426.73 & \multirow{2}{4em}{\centering 9.25} & \multirow{2}{4em}{\centering 9.25} & 1.119\\ 
& ${\mathrm{TE}}_{21,10,0}$ & 435.88 & 435.98 &  &  & 1.150\\ 
\hline
\end{tabular}
\end{center}
\end{table}

While for the case of Fig. \ref{Fig1} the identification of specific mode families based on their spectral fingerprint, (e.g. resonance full-width half maximum (FWHM) and intensity profile) is rather simple, it is clear from Fig. \ref{Fig2} that with the presence of an increased number of WGMs, experimental determination of mode families becomes very quickly cumbersome, if not impossible. A solution to readily identify corresponding mode families of a THz WGM resonator experimentally is by measuring the phase profile of the observed WGMs as a function of waveguide-resonator distance. By decreasing the waveguide-resonator distance (starting at a large waveguide-resonator distance) a specific WGM family will undergo a distinct phase change from under coupled, to critically coupled and over coupled. Over coupling and critical coupling show an overall phase change of 2$\pi$ centered around the resonance frequency. The spectral width of this 2$\pi$ phase transition decreases from a very broad transition 
for extreme over 
coupling to a step-function like transition for the case of critical coupling. Therefore, the phase information allows to evaluate how far a specific mode is detuned from critical coupling. The corresponding phase profiles and their characteristics are discussed in detail in our previous publications \cite{Vogt:172,Vogt:173}. 

Over a limited frequency range (here demonstrated for about 30\,GHz) all WGMs of one mode family will show a very similar transient of the phase profile upon a decrease in waveguide-resonator distance. In this small frequency range all WGMs of the same mode family will be over coupled before a WGM of another mode family will show over coupling. This can qualitatively be explained with the very similar phase-matching condition and overlap of the evanescent fields of the WGMs within one mode family. An example can be seen in Fig. \ref{Fig1} (b) where only the WGMs of one mode family at 0.411\,THz, 0.422\,THz and 0.432\,THz are over coupled while the other two mode families are under coupled. Upon further reduction of the waveguide-resonator distance also the second mode family (spectrally located to the right of the first mode family) will become over coupled (not shown here). The same behavior can be observed in Fig. \ref{Fig2} (b) where for a specific waveguide-resonator distance already two of the four main 
mode families are over coupled (apart from the WGM at 630.57\,GHz).

The technique described above allows to experimentally determine the FSRs of the corresponding WGMs and finally identify each WGM present in the spectrum based on a comparison with the numerical results. Table \ref{table1} and table \ref{table2} present a comparison of the measured resonance frequencies and FSRs with the results predicted by MDAK theory for the frequency ranges from 411\,GHz to 437\,GHz and 607\,GHz to 635\,GHz, respectively. The experimental results are in excellent agreement with the numerical values, and clearly allow an unambiguous identification of the WGMs present in the measured spectra. Please note that the resonance frequencies were measured for the case of very weak coupling (large waveguide-resonator distance) to avoid a red shift of the resonances due to the waveguide material, as the MDAK theory will give the frequencies for the unloaded WGM resonator \cite{Vogt:172}. The allocated WGMs are listed in the tables, but are also partially labeled in Fig. \ref{Fig1} (a) and Fig. \ref{Fig2} (a) for convenience. The first index $p$ of the 
nomenclature ${\mathrm{TE}}_{p,q,l}$ states the number of wavelengths in the 
circumference of the spherical resonator. While $q$ counts the number of nodes in radial direction ($q>\,0$ are referred to as higher order radial modes), and $l$ accounts for higher order azimuthal modes (number of nodes from pole to pole, perpendicular to the equatorial plane). Please note that higher order azimuthal modes ($l>\,0$) are degenerate for a perfect sphere \cite{matsko2006optical}.

All experimentally observed WGMs are higher order radial modes. For example the ${\mathrm{TE}}_{\mathrm{36,12,0}}$ at 618.15\,GHz has 36 wavelength in the circumference of the resonator and 12 nodes in the radial direction. The radial energy density distribution of this WGM is shown in Fig. \ref{Fig3} (c). The explanation why particularly those WGMs are observed is discussed in detail in section \ref{phasematching} phase matching.

\begin{table}[t]
\caption{Comparison of the experimentally obtained resonance frequencies (${\mathrm{f}}_{\mathrm{exp}}$) and FSRs (${\mathrm{FSR}}_{\mathrm{exp}}$) with the numerical results from the MDAK theory (${\mathrm{f}}_{\mathrm{MDAK}}$, ${\mathrm{FSR}}_\mathrm{{MDAK}}$) in the frequency range 607\,GHz to 635\,GHz. The rows are clustered according to their radial order, and the corresponding WGMs are labeled in the second column. The last column shows the calculated effective refractive index of the WGMs at the surface of the resonator (${\mathrm{n}}_{\mathrm{ph}}$\raisebox{-.2em}{$\vert_{{r}_{0}}$}).\label{table2} }
\begin{center}
\begin{tabular}{ |c|c|c|c|c|c|c| } 
\hline
\makecell{radial order\\ $q$}& Mode & ${\mathrm{f}}_{\mathrm{exp}} [GHZ]$ & ${\mathrm{f}}_{\mathrm{MDAK}} [GHZ]$  & \makecell{${\mathrm{FSR}}_{\mathrm{exp}}$\\ $[GHZ]$} & \makecell{${\mathrm{FSR}}_{\mathrm{MDAK}}$\\ $[GHZ]$} & ${\mathrm{n}}_{\mathrm{ph}}$\raisebox{-.2em}{$\big\vert_{{r}_{0}}$} \\
\hline
\multirow{3}{4em}{\centering 10} & ${\mathrm{TE}}_{41,10,0}$ & 613.24 & 613.39 & \multirow{2}{4em}{\centering 8.60} & \multirow{2}{4em}{\centering 8.60} & 1.596\\ 
& ${\mathrm{TE}}_{42,10,0}$ & 621.84 & 621.99 & \multirow{2}{4em}{\centering 8.59} & \multirow{2}{4em}{\centering 8.58} & 1.612\\ 
& ${\mathrm{TE}}_{43,10,0}$ & 630.43 & 630.57 &  &  & 1.628\\ 
\hline
\multirow{3}{4em}{\centering 11} & ${\mathrm{TE}}_{38,11,0}$ & 611.68 & 611.83 & \multirow{2}{4em}{\centering 8.74} & \multirow{2}{4em}{\centering 8.73} & 1.483\\ 
& ${\mathrm{TE}}_{39,11,0}$ & 620.41 & 620.56 & \multirow{2}{4em}{\centering 8.72} & \multirow{2}{4em}{\centering 8.71} & 1.500\\ 
& ${\mathrm{TE}}_{40,11,0}$ & 629.13 & 629.27 &  &  & 1.518\\ 
\hline
\multirow{3}{4em}{\centering 12} & ${\mathrm{TE}}_{35,12,0}$ & 609.28 & 609.43 & \multirow{2}{4em}{\centering 8.88} & \multirow{2}{4em}{\centering 8.87} & 1.371\\ 
& ${\mathrm{TE}}_{36,12,0}$ & 618.15 & 618.30 & \multirow{2}{4em}{\centering 8.85} & \multirow{2}{4em}{\centering 8.85} & 1.390\\ 
& ${\mathrm{TE}}_{37,12,0}$ & 627.00 & 627.14 &  &  & 1.408\\ 
\hline
\multirow{3}{4em}{\centering 13} & ${\mathrm{TE}}_{33,13,0}$ & 615.09 & 615.23 & \multirow{2}{4em}{\centering 8.99} & \multirow{2}{4em}{\centering 8.99} & 1.281\\ 
& ${\mathrm{TE}}_{34,13,0}$ & 624.08 & 624.22 & \multirow{2}{4em}{\centering 8.97} & \multirow{2}{4em}{\centering 8.96} & 1.300\\ 
& ${\mathrm{TE}}_{35,13,0}$ & 633.05 & 633.18 &  &  & 1.320\\ 
\hline
\end{tabular}
\end{center}
\end{table}

\section{Quality factor}

The Q factor, defined as the resonance frequency divided by the FWHM of the resonance, is an essential figure of merit for any WGM resonator. A narrow resonance is desirable for various applications based on WGMs. As discussed above the main limitation for the Q factor of THz WGMs is the absorption of the resonator material. 

\begin{figure}[t]
\centering
\includegraphics[width=13.5cm]{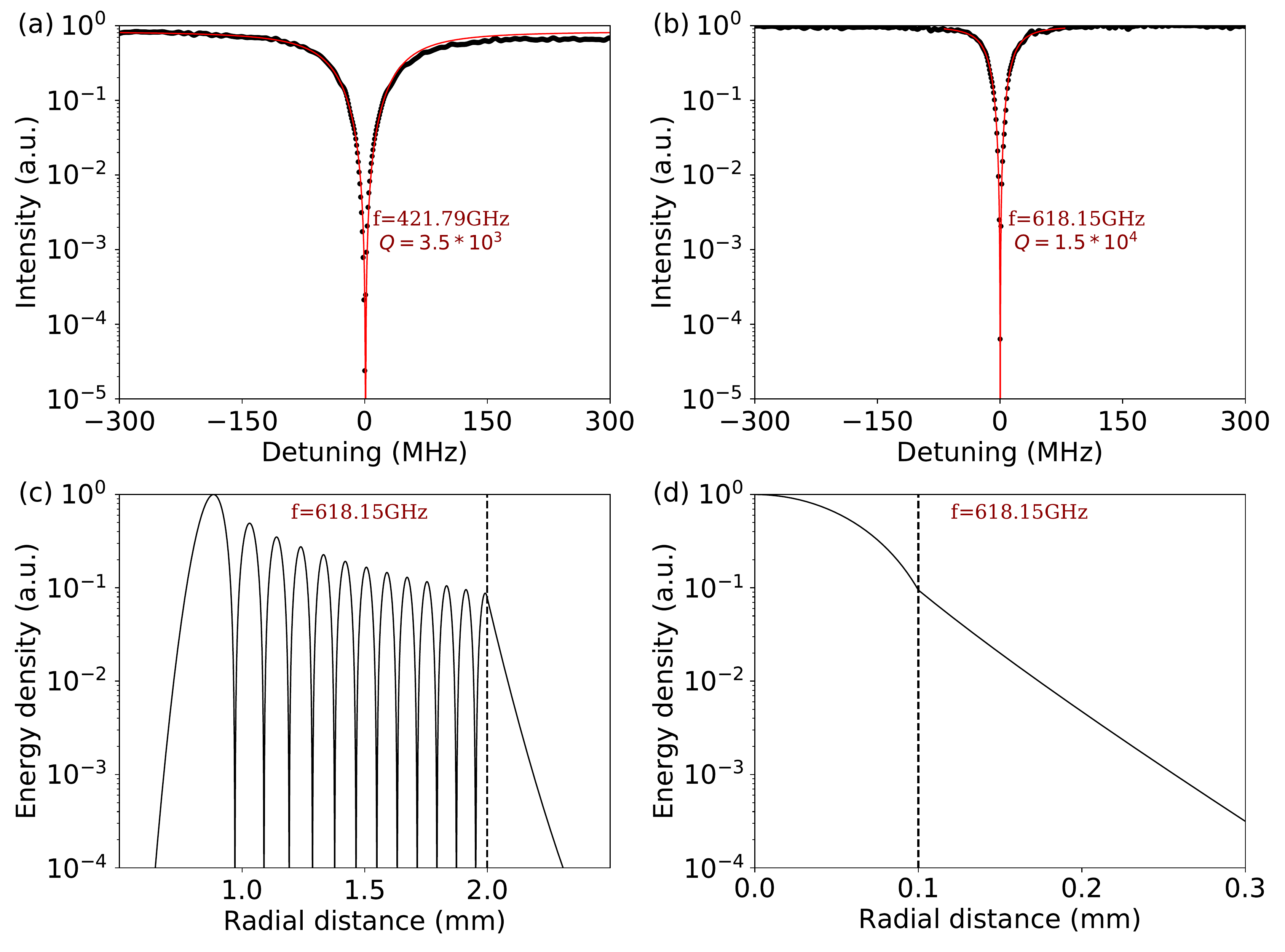}%
\caption{(a) and (b) show the normalized measured intensity transmission (black dots) of the ${\mathrm{TE}}_{\mathrm{22,9,0}}$ WGM and ${\mathrm{TE}}_{\mathrm{36,12,0}}$ WGM at critical coupling on a logarithmic scale, respectively. The red line shows the fitted Lorentzian function. The Q factor extracted from the fit of the ${\mathrm{TE}}_{\mathrm{22,9,0}}$ WGM is $3.5 \times {10}^{3}$ and $1.5 \times {10}^{4}$ for the ${\mathrm{TE}}_{\mathrm{36,12,0}}$ WGM, and (c) depicts the normalized energy density distribution in radial direction of the ${\mathrm{TE}}_{\mathrm{36,12,0}}$ WGM calculated with MDAK theory on a logarithmic scale. Please note, that the shown energy density is averaged over the polar and azimuthal angle. The black dashed line indicates the radius of the HRFZ-Si sphere, and (d) calculated normalized energy density distribution in radial direction of the air-silica step index waveguide at 618.15\,GHz. The black dashed line indicates the radius of the sub-wavelength waveguide.\label{Fig3}}%
\end{figure}

Fig. \ref{Fig3} (a) and (b) show the measured normalized intensity profiles of the ${\mathrm{TE}}_{\mathrm{22,9,0}}$ WGM and ${\mathrm{TE}}_{\mathrm{36,12,0}}$ WGM at 421.79\,GHz and 618.15\,GHz, respectively, at critical coupling. The scans are averaged over three independent measurements and the frequency step size is 1\,MHz. Both WGMs show ultra high Q factors of $3.5 \times {10}^{3}$ (${\mathrm{TE}}_{\mathrm{22,9,0}}$ WGM) and $1.5 \times {10}^{4}$ (${\mathrm{TE}}_{\mathrm{36,12,0}}$ WGM). To the best of our knowledge these are by far the highest Q factors ever reported for a THz WGM resonator. The FWHM of the ${\mathrm{TE}}_{\mathrm{36,12,0}}$ WGM is about 41\,MHz, and the finesse is about 220. 

Interestingly, the extracted Q factor of the two WGMs differ by a factor of about 4.3. If both WGMs would experience the same losses, the Q factor would only differ by a factor of about 1.5, simply due to the higher resonance frequency. Assuming a material absorption of $0.015 {\mathrm{cm}}^{-1}$ of the HRFZ-Si in the MDAK calculations reproduces the Q factor of the ${\mathrm{TE}}_{\mathrm{36,12,0}}$ WGM of $1.5 \times {10}^{4}$. At the same time, the MDAK calculations show an about 3.5 times lower Q factor of the ${\mathrm{TE}}_{\mathrm{22,9,0}}$ WGM. The discrepancy in the drop of the Q factor between the measurements (by a factor of about 4.3) and the numerical analysis (by a factor of about 3.5) can only be accounted for by a frequency dependent material absorption. Introducing a 35\,$\%$ higher material absorption of $0.023 {\mathrm{cm}}^{-1}$ around 420\,GHz in the MDAK calculations reproduces the Q factor observed in the measurement. Considering the large uncertainty for the material absorption of 
HRFZ-
Si reported in the literature using THZ time-domain spectroscopy, in particular below 0.5\,THz, the observed frequency dependent material absorption is well within the expectations \cite{naftaly2016international}. Furthermore, the decrease in the Q factor by a factor of 3.5 from MDAK theory compared to the expected factor of 1.5 can be explained with higher radiation losses of the ${\mathrm{TE}}_{\mathrm{22,9,0}}$ WGM at about 420\,GHz compared to the ${\mathrm{TE}}_{\mathrm{36,12,0}}$ WGM at about 620\,GHz. The higher radiation losses, similar to bend losses for a waveguide, observed in the 0.4\,THz frequency range compared to the 0.6\,THz frequency range can qualitatively be explained with the larger ratio of the wavelength to the radius of the resonator.

Finally, according to the computations based on MDAK theory, the experimentally observed higher order radial WGMs from 607\,GHz to 635\,GHz have exactly the same Q factor as the fundamental WGM ($q,l=0$) in the same frequency range. This is in contrast to the bend loss dominated WGMs in the frequency range from 411\,GHz to 437\,GHz, where numerically a significant difference in the Q factor between the fundamental mode and the higher order radial modes are observed. The impact of the bend loss in this frequency range is also qualitatively evident in the experimental data presented in Fig. \ref{Fig1}: the 10th higher order radial mode family has already a significant wider FWHM compared to the 9th higher order radial mode family. This observation is in excellent agreement with the numerical predictions.

\section{Phase matching}
\label{phasematching}

In order to achieve strong coupling of a WGM via the evanescent field from a waveguide mode, the modes need to have an overlap of the evanescent fields and similar phase refractive indices \cite{matsko2006optical}.         

The phase refractive indices ${\mathrm{n}}_{\mathrm{ph}}$ of the identified WGMs at the surface of the sphere ($r={r}_{0}$) can in first order be calculated with the following equation: ${\mathrm{n}}_{\mathrm{ph}}$\raisebox{-.2em}{$\vert_{{r}_{0}}$}$=(pc)/(2\pi {r}_{0}f)$ \cite{Foreman:15}. With $p$ the number of wavelength in the circumference of the resonator, $c$ the speed of light, ${r}_{0}$ the resonator radius, and $f$ the unloaded resonance frequency of the WGM. The equation is the condition for constructive interference of the mode after propagation around the resonator circumference.

The phase refractive indices of the experimentally observed WGMs calculated with the equation above are summarized in table \ref{table1} and table \ref{table2}, respectively. These WGMs in the frequency range from 411\,GHz to 437\,GHz have phase refractive indices in the range from 1.1 to 1.4, while the WGMs between 607\,GHz to 635\,GHz show ${\mathrm{n}}_{\mathrm{ph}}$\raisebox{-.2em}{$\vert_{{r}_{0}}$} of 1.3 to 1.6. For comparison, the calculated phase refractive indices of the waveguide mode are between 1.03 to 1.06 in the frequency range from 411\,GHz to 437\,GHz, and between 1.32 to 1.35 from 607\,GHz to 635\,GHz.

While the calculated phase mismatches between the WGMs and the waveguide mode are not insignificant, the observed critical coupling of those WGMs with the sub-wavelength silica waveguide is well within the expectations. On one hand, the short interaction length between the WGM and the waveguide mode prohibits the agglomeration of a significant phase difference between the interacting modes \cite{Ristic:13}; on the other hand, due to the large wavelength compared to our structures, the waveguide mode and the WGMs show a vast overlap of the evanescent fields (see Fig. \ref{Fig3} (c) and (d)). The latter facilitates critical coupling of the WGM despite the relatively large phase mismatch \cite{Ristic:13}.

Consequently, the simple approximation of the phase refractive index of the higher order radial WGMs at the surface allows to predict the observed coupling between the HRFZ-Si THz WGM resonator and the sub-wavelength silica waveguide. However, due to the large mode volume of the THz WGMs, the calculated ${\mathrm{n}}_{\mathrm{ph}}$\raisebox{-.2em}{$\vert_{{r}_{0}}$} does not constitute the effective refractive indices of the WGMs that is usually assigned to a mode but is rather to be understood as a localized phase refractive index as a function of the distance from the centre of the sphere. The choice to use the radius of the resonator in the above equation is motivated by the fact that the interaction between the WGM and the waveguide mode is localized in the vicinity of the resonator surface. Therefore the calculated ${\mathrm{n}}_{\mathrm{ph}}$\raisebox{-.2em}{$\vert_{{r}_{0}}$} provides a reasonable estimate for the phase refractive indices of the higher order radial THz WGMs essential for the 
evanescent coupling, and allows to qualitatively explain the experimental results.

\section{Conclusion}

The presented results demonstrate the first experimental observation of critically coupled ultra high Q higher order radial THz WGMs. Critical coupling is achieved despite the significant discrepancy in the refractive indices of the two materials used for the resonator and the waveguide, namely silicon and silica.

The observed WGM spectra can be fully explained using just the phase profiles from the coherent CW THz spectroscopy measurements and the numerical results from MDAK theory. The experimental and numerical results show excellent agreement allowing to unambiguously identify the WGMs present in the measured spectra. The higher order radial modes have an extremely high Q factor of $1.5 \times {10}^{4}$ at 0.62\,THz extracted from the measured intensity profiles. Furthermore, a simple approximation of the localized phase refractive index of the higher radial order THz WGMs at the surface of the resonator enables to qualitatively explain the observed coupling between the HRFZ-Si sphere and a silica waveguide. The obtained results highlight the extenuated phase matching criteria for the observed THz WGMs, but also provides a powerful tool to engineer the resonator/waveguide material combination for the excitation of specific THz WGM modes.

We believe that the presented results provide vast opportunities for the establishment of technologies and applications based on WGM resonators with ultra high Q factors in the THz frequency range.

%

%


\begin{thebibliography}{34}%
\makeatletter
\providecommand \@ifxundefined [1]{%
 \@ifx{#1\undefined}
}%
\providecommand \@ifnum [1]{%
 \ifnum #1\expandafter \@firstoftwo
 \else \expandafter \@secondoftwo
 \fi
}%
\providecommand \@ifx [1]{%
 \ifx #1\expandafter \@firstoftwo
 \else \expandafter \@secondoftwo
 \fi
}%
\providecommand \natexlab [1]{#1}%
\providecommand \enquote  [1]{``#1''}%
\providecommand \bibnamefont  [1]{#1}%
\providecommand \bibfnamefont [1]{#1}%
\providecommand \citenamefont [1]{#1}%
\providecommand \href@noop [0]{\@secondoftwo}%
\providecommand \href [0]{\begingroup \@sanitize@url \@href}%
\providecommand \@href[1]{\@@startlink{#1}\@@href}%
\providecommand \@@href[1]{\endgroup#1\@@endlink}%
\providecommand \@sanitize@url [0]{\catcode `\\12\catcode `\$12\catcode
  `\&12\catcode `\#12\catcode `\^12\catcode `\_12\catcode `\%12\relax}%
\providecommand \@@startlink[1]{}%
\providecommand \@@endlink[0]{}%
\providecommand \url  [0]{\begingroup\@sanitize@url \@url }%
\providecommand \@url [1]{\endgroup\@href {#1}{\urlprefix }}%
\providecommand \urlprefix  [0]{URL }%
\providecommand \Eprint [0]{\href }%
\providecommand \doibase [0]{http://dx.doi.org/}%
\providecommand \selectlanguage [0]{\@gobble}%
\providecommand \bibinfo  [0]{\@secondoftwo}%
\providecommand \bibfield  [0]{\@secondoftwo}%
\providecommand \translation [1]{[#1]}%
\providecommand \BibitemOpen [0]{}%
\providecommand \bibitemStop [0]{}%
\providecommand \bibitemNoStop [0]{.\EOS\space}%
\providecommand \EOS [0]{\spacefactor3000\relax}%
\providecommand \BibitemShut  [1]{\csname bibitem#1\endcsname}%
\let\auto@bib@innerbib\@empty
\bibitem [{\citenamefont {Matsko}\ and\ \citenamefont
  {Ilchenko}(2006)}]{matsko2006optical}%
  \BibitemOpen
  \bibfield  {author} {\bibinfo {author} {\bibfnamefont {A.~B.}\ \bibnamefont
  {Matsko}}\ and\ \bibinfo {author} {\bibfnamefont {V.~S.}\ \bibnamefont
  {Ilchenko}},\ }\href@noop {} {\bibfield  {journal} {\bibinfo  {journal} {IEEE
  J. Sel. Top. Quantum Electron}\ }\textbf {\bibinfo {volume} {12}},\ \bibinfo
  {pages} {3} (\bibinfo {year} {2006})}\BibitemShut {NoStop}%
\bibitem [{\citenamefont {Chiasera}\ \emph {et~al.}(2010)\citenamefont
  {Chiasera}, \citenamefont {Dumeige}, \citenamefont {Féron}, \citenamefont
  {Ferrari}, \citenamefont {Jestin}, \citenamefont {Nunzi~Conti}, \citenamefont
  {Pelli}, \citenamefont {Soria},\ and\ \citenamefont
  {Righini}}]{LPOR:LPOR200910016}%
  \BibitemOpen
  \bibfield  {author} {\bibinfo {author} {\bibfnamefont {A.}~\bibnamefont
  {Chiasera}}, \bibinfo {author} {\bibfnamefont {Y.}~\bibnamefont {Dumeige}},
  \bibinfo {author} {\bibfnamefont {P.}~\bibnamefont {Féron}}, \bibinfo
  {author} {\bibfnamefont {M.}~\bibnamefont {Ferrari}}, \bibinfo {author}
  {\bibfnamefont {Y.}~\bibnamefont {Jestin}}, \bibinfo {author} {\bibfnamefont
  {G.}~\bibnamefont {Nunzi~Conti}}, \bibinfo {author} {\bibfnamefont
  {S.}~\bibnamefont {Pelli}}, \bibinfo {author} {\bibfnamefont
  {S.}~\bibnamefont {Soria}}, \ and\ \bibinfo {author} {\bibfnamefont
  {G.}~\bibnamefont {Righini}},\ }\href {\doibase 10.1002/lpor.200910016}
  {\bibfield  {journal} {\bibinfo  {journal} {Laser and Photonics Reviews}\
  }\textbf {\bibinfo {volume} {4}},\ \bibinfo {pages} {457} (\bibinfo {year}
  {2010})}\BibitemShut {NoStop}%
\bibitem [{\citenamefont {Foreman}\ \emph {et~al.}(2015)\citenamefont
  {Foreman}, \citenamefont {Swaim},\ and\ \citenamefont
  {Vollmer}}]{Foreman:15}%
  \BibitemOpen
  \bibfield  {author} {\bibinfo {author} {\bibfnamefont {M.~R.}\ \bibnamefont
  {Foreman}}, \bibinfo {author} {\bibfnamefont {J.~D.}\ \bibnamefont {Swaim}},
  \ and\ \bibinfo {author} {\bibfnamefont {F.}~\bibnamefont {Vollmer}},\ }\href
  {\doibase 10.1364/AOP.7.000168} {\bibfield  {journal} {\bibinfo  {journal}
  {Adv. Opt. Photon.}\ }\textbf {\bibinfo {volume} {7}},\ \bibinfo {pages}
  {168} (\bibinfo {year} {2015})}\BibitemShut {NoStop}%
\bibitem [{\citenamefont {Ilchenko}\ and\ \citenamefont
  {Matsko}(2006)}]{ilchenko2006optical}%
  \BibitemOpen
  \bibfield  {author} {\bibinfo {author} {\bibfnamefont {V.~S.}\ \bibnamefont
  {Ilchenko}}\ and\ \bibinfo {author} {\bibfnamefont {A.~B.}\ \bibnamefont
  {Matsko}},\ }\href@noop {} {\bibfield  {journal} {\bibinfo  {journal} {IEEE
  Journal of selected topics in quantum electronics}\ }\textbf {\bibinfo
  {volume} {12}},\ \bibinfo {pages} {15} (\bibinfo {year} {2006})}\BibitemShut
  {NoStop}%
\bibitem [{\citenamefont {Yang}\ \emph {et~al.}(2014)\citenamefont {Yang},
  \citenamefont {Ward},\ and\ \citenamefont {Chormaic}}]{yang2014quasi}%
  \BibitemOpen
  \bibfield  {author} {\bibinfo {author} {\bibfnamefont {Y.}~\bibnamefont
  {Yang}}, \bibinfo {author} {\bibfnamefont {J.}~\bibnamefont {Ward}}, \ and\
  \bibinfo {author} {\bibfnamefont {S.~N.}\ \bibnamefont {Chormaic}},\
  }\href@noop {} {\bibfield  {journal} {\bibinfo  {journal} {Optics express}\
  }\textbf {\bibinfo {volume} {22}},\ \bibinfo {pages} {6881} (\bibinfo {year}
  {2014})}\BibitemShut {NoStop}%
\bibitem [{\citenamefont {Naftaly}\ and\ \citenamefont
  {Miles}(2007)}]{naftaly2007terahertz}%
  \BibitemOpen
  \bibfield  {author} {\bibinfo {author} {\bibfnamefont {M.}~\bibnamefont
  {Naftaly}}\ and\ \bibinfo {author} {\bibfnamefont {R.~E.}\ \bibnamefont
  {Miles}},\ }\href@noop {} {\bibfield  {journal} {\bibinfo  {journal}
  {Proceedings of the IEEE}\ }\textbf {\bibinfo {volume} {95}},\ \bibinfo
  {pages} {1658} (\bibinfo {year} {2007})}\BibitemShut {NoStop}%
\bibitem [{\citenamefont {Annino}\ \emph {et~al.}(1997)\citenamefont {Annino},
  \citenamefont {Cassettari}, \citenamefont {Longo},\ and\ \citenamefont
  {Martinelli}}]{annino1997whispering}%
  \BibitemOpen
  \bibfield  {author} {\bibinfo {author} {\bibfnamefont {G.}~\bibnamefont
  {Annino}}, \bibinfo {author} {\bibfnamefont {M.}~\bibnamefont {Cassettari}},
  \bibinfo {author} {\bibfnamefont {I.}~\bibnamefont {Longo}}, \ and\ \bibinfo
  {author} {\bibfnamefont {M.}~\bibnamefont {Martinelli}},\ }\href@noop {}
  {\bibfield  {journal} {\bibinfo  {journal} {IEEE Transactions on microwave
  theory and techniques}\ }\textbf {\bibinfo {volume} {45}},\ \bibinfo {pages}
  {2025} (\bibinfo {year} {1997})}\BibitemShut {NoStop}%
\bibitem [{\citenamefont {Preu}\ \emph {et~al.}(2013)\citenamefont {Preu},
  \citenamefont {Schmid}, \citenamefont {Sedlmeir}, \citenamefont {Evers},\
  and\ \citenamefont {Schwefel}}]{Preu13}%
  \BibitemOpen
  \bibfield  {author} {\bibinfo {author} {\bibfnamefont {S.}~\bibnamefont
  {Preu}}, \bibinfo {author} {\bibfnamefont {S.~I.}\ \bibnamefont {Schmid}},
  \bibinfo {author} {\bibfnamefont {F.}~\bibnamefont {Sedlmeir}}, \bibinfo
  {author} {\bibfnamefont {J.}~\bibnamefont {Evers}}, \ and\ \bibinfo {author}
  {\bibfnamefont {H.~G.}\ \bibnamefont {Schwefel}},\ }\href@noop {} {\bibfield
  {journal} {\bibinfo  {journal} {Optics express}\ }\textbf {\bibinfo {volume}
  {21}},\ \bibinfo {pages} {16370} (\bibinfo {year} {2013})}\BibitemShut
  {NoStop}%
\bibitem [{\citenamefont {Preu}\ \emph {et~al.}(2008)\citenamefont {Preu},
  \citenamefont {Schwefel}, \citenamefont {Malzer}, \citenamefont {D\"{o}hler},
  \citenamefont {Wang}, \citenamefont {Hanson}, \citenamefont {Zimmerman},\
  and\ \citenamefont {Gossard}}]{Preu:08}%
  \BibitemOpen
  \bibfield  {author} {\bibinfo {author} {\bibfnamefont {S.}~\bibnamefont
  {Preu}}, \bibinfo {author} {\bibfnamefont {H.~G.~L.}\ \bibnamefont
  {Schwefel}}, \bibinfo {author} {\bibfnamefont {S.}~\bibnamefont {Malzer}},
  \bibinfo {author} {\bibfnamefont {G.~H.}\ \bibnamefont {D\"{o}hler}},
  \bibinfo {author} {\bibfnamefont {L.~J.}\ \bibnamefont {Wang}}, \bibinfo
  {author} {\bibfnamefont {M.}~\bibnamefont {Hanson}}, \bibinfo {author}
  {\bibfnamefont {J.~D.}\ \bibnamefont {Zimmerman}}, \ and\ \bibinfo {author}
  {\bibfnamefont {A.~C.}\ \bibnamefont {Gossard}},\ }\href {\doibase
  10.1364/OE.16.007336} {\bibfield  {journal} {\bibinfo  {journal} {Opt.
  Express}\ }\textbf {\bibinfo {volume} {16}},\ \bibinfo {pages} {7336}
  (\bibinfo {year} {2008})}\BibitemShut {NoStop}%
\bibitem [{\citenamefont {Vogt}\ and\ \citenamefont
  {Leonhardt}(2017{\natexlab{a}})}]{Vogt:172}%
  \BibitemOpen
  \bibfield  {author} {\bibinfo {author} {\bibfnamefont {D.~W.}\ \bibnamefont
  {Vogt}}\ and\ \bibinfo {author} {\bibfnamefont {R.}~\bibnamefont
  {Leonhardt}},\ }\href {\doibase 10.1364/OPTICA.4.000809} {\bibfield
  {journal} {\bibinfo  {journal} {Optica}\ }\textbf {\bibinfo {volume} {4}},\
  \bibinfo {pages} {809} (\bibinfo {year} {2017}{\natexlab{a}})}\BibitemShut
  {NoStop}%
\bibitem [{\citenamefont {Zhang}\ and\ \citenamefont
  {Grischkowsky}(2003)}]{Zhang:03}%
  \BibitemOpen
  \bibfield  {author} {\bibinfo {author} {\bibfnamefont {J.}~\bibnamefont
  {Zhang}}\ and\ \bibinfo {author} {\bibfnamefont {D.}~\bibnamefont
  {Grischkowsky}},\ }\href {\doibase 10.1364/JOSAB.20.001894} {\bibfield
  {journal} {\bibinfo  {journal} {J. Opt. Soc. Am. B}\ }\textbf {\bibinfo
  {volume} {20}},\ \bibinfo {pages} {1894} (\bibinfo {year}
  {2003})}\BibitemShut {NoStop}%
\bibitem [{\citenamefont {Vogt}\ and\ \citenamefont
  {Leonhardt}(2017{\natexlab{c}})}]{Vogt:173}%
  \BibitemOpen
  \bibfield  {author} {\bibinfo {author} {\bibfnamefont {D.~W.}\ \bibnamefont
  {Vogt}}\ and\ \bibinfo {author} {\bibfnamefont {R.}~\bibnamefont
  {Leonhardt}},\ }\href {\doibase 10.1364/OL.42.004359} {\bibfield  {journal}
  {\bibinfo  {journal} {Opt. Lett.}\ }\textbf {\bibinfo {volume} {42}},\
  \bibinfo {pages} {4359} (\bibinfo {year} {2017}{\natexlab{c}})}\BibitemShut
  {NoStop}%
\bibitem [{\citenamefont {Nagel}\ \emph {et~al.}(2006)\citenamefont {Nagel},
  \citenamefont {F{\"o}rst},\ and\ \citenamefont {Kurz}}]{nagel2006thz}%
  \BibitemOpen
  \bibfield  {author} {\bibinfo {author} {\bibfnamefont {M.}\ \bibnamefont
  {Nagel}}, \bibinfo {author} {\bibfnamefont {M.}~\bibnamefont {F{\"o}rst}}, \
  and\ \bibinfo {author} {\bibfnamefont {H.}~\bibnamefont {Kurz}},\
  }\href@noop {} {\bibfield  {journal} {\bibinfo  {journal} {Journal of Physics: Condensed Matter}\ }\textbf {\bibinfo {volume} {18}},\ \bibinfo {pages} {S601}
  (\bibinfo {year} {2006})}\BibitemShut {NoStop}%
\bibitem [{\citenamefont {Cao}\ \emph {et~al.}(2012)\citenamefont {Cao},
  \citenamefont {Singh}, \citenamefont {Al-Naib}, \citenamefont {He},
  \citenamefont {Taylor},\ and\ \citenamefont {Zhang}}]{cao2012low}%
  \BibitemOpen
  \bibfield  {author} {\bibinfo {author} {\bibfnamefont {W.}~\bibnamefont
  {Cao}}, \bibinfo {author} {\bibfnamefont {R.}~\bibnamefont {Singh}}, \bibinfo
  {author} {\bibfnamefont {I.~A.}\ \bibnamefont {Al-Naib}}, \bibinfo {author}
  {\bibfnamefont {M.}~\bibnamefont {He}}, \bibinfo {author} {\bibfnamefont
  {A.~J.}\ \bibnamefont {Taylor}}, \ and\ \bibinfo {author} {\bibfnamefont
  {W.}~\bibnamefont {Zhang}},\ }\href@noop {} {\bibfield  {journal} {\bibinfo
  {journal} {Optics letters}\ }\textbf {\bibinfo {volume} {37}},\ \bibinfo
  {pages} {3366} (\bibinfo {year} {2012})}\BibitemShut {NoStop}%
\bibitem [{\citenamefont {Yee}\ and\ \citenamefont
  {Sherwin}(2009)}]{yee2009high}%
  \BibitemOpen
  \bibfield  {author} {\bibinfo {author} {\bibfnamefont {C.~M.}\ \bibnamefont
  {Yee}}\ and\ \bibinfo {author} {\bibfnamefont {M.~S.}\ \bibnamefont
  {Sherwin}},\ }\href@noop {} {\bibfield  {journal} {\bibinfo  {journal}
  {Applied Physics Letters}\ }\textbf {\bibinfo {volume} {94}},\ \bibinfo
  {pages} {154104} (\bibinfo {year} {2009})}\BibitemShut {NoStop}%
\bibitem [{\citenamefont {Li}\ \emph {et~al.}(2017)\citenamefont {Li},
  \citenamefont {Nallappan}, \citenamefont {Guerboukha},\ and\ \citenamefont
  {Skorobogatiy}}]{Li:17}%
  \BibitemOpen
  \bibfield  {author} {\bibinfo {author} {\bibfnamefont {J.}~\bibnamefont
  {Li}}, \bibinfo {author} {\bibfnamefont {K.}~\bibnamefont {Nallappan}},
  \bibinfo {author} {\bibfnamefont {H.}~\bibnamefont {Guerboukha}}, \ and\
  \bibinfo {author} {\bibfnamefont {M.}~\bibnamefont {Skorobogatiy}},\ }\href
  {\doibase 10.1364/OE.25.004126} {\bibfield  {journal} {\bibinfo  {journal}
  {Opt. Express}\ }\textbf {\bibinfo {volume} {25}},\ \bibinfo {pages} {4126}
  (\bibinfo {year} {2017})}\BibitemShut {NoStop}%
\bibitem [{\citenamefont {Mendis}\ and\ \citenamefont
  {Mittleman}(2009)}]{mendis2009comparison}%
  \BibitemOpen
  \bibfield  {author} {\bibinfo {author} {\bibfnamefont {R.}~\bibnamefont
  {Mendis}}\ and\ \bibinfo {author} {\bibfnamefont {D.~M.}\ \bibnamefont
  {Mittleman}},\ }\href@noop {} {\bibfield  {journal} {\bibinfo  {journal}
  {Optics express}\ }\textbf {\bibinfo {volume} {17}},\ \bibinfo {pages}
  {14839} (\bibinfo {year} {2009})}\BibitemShut {NoStop}%
\bibitem [{\citenamefont {Bingham}\ and\ \citenamefont
  {Grischkowsky}(2008)}]{bingham2008terahertz}%
  \BibitemOpen
  \bibfield  {author} {\bibinfo {author} {\bibfnamefont {A.}~\bibnamefont
  {Bingham}}\ and\ \bibinfo {author} {\bibfnamefont {D.}~\bibnamefont
  {Grischkowsky}},\ }\href@noop {} {\bibfield  {journal} {\bibinfo  {journal}
  {Optics letters}\ }\textbf {\bibinfo {volume} {33}},\ \bibinfo {pages} {348}
  (\bibinfo {year} {2008})}\BibitemShut {NoStop}%
\bibitem [{\citenamefont {Mendis}\ \emph {et~al.}(2009)\citenamefont {Mendis},
  \citenamefont {Astley}, \citenamefont {Liu},\ and\ \citenamefont
  {Mittleman}}]{mendis2009terahertz}%
  \BibitemOpen
  \bibfield  {author} {\bibinfo {author} {\bibfnamefont {R.}~\bibnamefont
  {Mendis}}, \bibinfo {author} {\bibfnamefont {V.}~\bibnamefont {Astley}},
  \bibinfo {author} {\bibfnamefont {J.}~\bibnamefont {Liu}}, \ and\ \bibinfo
  {author} {\bibfnamefont {D.~M.}\ \bibnamefont {Mittleman}},\ }\href@noop {}
  {\bibfield  {journal} {\bibinfo  {journal} {Applied Physics Letters}\
  }\textbf {\bibinfo {volume} {95}},\ \bibinfo {pages} {171113} (\bibinfo
  {year} {2009})}\BibitemShut {NoStop}%
\bibitem [{\citenamefont {George}\ \emph {et~al.}(2007)\citenamefont {George},
  \citenamefont {Manolatou}, \citenamefont {Rana}, \citenamefont {Bingham},\
  and\ \citenamefont {Grischkowsky}}]{george2007integrated}%
  \BibitemOpen
  \bibfield  {author} {\bibinfo {author} {\bibfnamefont {P.~A.}\ \bibnamefont
  {George}}, \bibinfo {author} {\bibfnamefont {C.}~\bibnamefont {Manolatou}},
  \bibinfo {author} {\bibfnamefont {F.}~\bibnamefont {Rana}}, \bibinfo {author}
  {\bibfnamefont {A.~L.}\ \bibnamefont {Bingham}}, \ and\ \bibinfo {author}
  {\bibfnamefont {D.~R.}\ \bibnamefont {Grischkowsky}},\ }\href@noop {}
  {\bibfield  {journal} {\bibinfo  {journal} {Applied Physics Letters}\
  }\textbf {\bibinfo {volume} {91}},\ \bibinfo {pages} {191122} (\bibinfo
  {year} {2007})}\BibitemShut {NoStop}%
\bibitem [{\citenamefont {Xu}\ \emph {et~al.}(2016)\citenamefont {Xu},
  \citenamefont {Singh},\ and\ \citenamefont {Zhang}}]{xu2016high}%
  \BibitemOpen
  \bibfield  {author} {\bibinfo {author} {\bibfnamefont {N.}~\bibnamefont
  {Xu}}, \bibinfo {author} {\bibfnamefont {R.}~\bibnamefont {Singh}}, \ and\
  \bibinfo {author} {\bibfnamefont {W.}~\bibnamefont {Zhang}},\ }\href@noop {}
  {\bibfield  {journal} {\bibinfo  {journal} {Applied Physics Letters}\
  }\textbf {\bibinfo {volume} {109}},\ \bibinfo {pages} {021108} (\bibinfo
  {year} {2016})}\BibitemShut {NoStop}%
\bibitem [{\citenamefont {Yang}\ \emph {et~al.}(2017)\citenamefont {Yang},
  \citenamefont {Tang}, \citenamefont {Liu}, \citenamefont {Wang},
  \citenamefont {Wang}, \citenamefont {Li}, \citenamefont {Wang},\ and\
  \citenamefont {Gu}}]{Yang:17}%
  \BibitemOpen
  \bibfield  {author} {\bibinfo {author} {\bibfnamefont {S.}~\bibnamefont
  {Yang}}, \bibinfo {author} {\bibfnamefont {C.}~\bibnamefont {Tang}}, \bibinfo
  {author} {\bibfnamefont {Z.}~\bibnamefont {Liu}}, \bibinfo {author}
  {\bibfnamefont {B.}~\bibnamefont {Wang}}, \bibinfo {author} {\bibfnamefont
  {C.}~\bibnamefont {Wang}}, \bibinfo {author} {\bibfnamefont {J.}~\bibnamefont
  {Li}}, \bibinfo {author} {\bibfnamefont {L.}~\bibnamefont {Wang}}, \ and\
  \bibinfo {author} {\bibfnamefont {C.}~\bibnamefont {Gu}},\ }\href {\doibase
  10.1364/OE.25.015938} {\bibfield  {journal} {\bibinfo  {journal} {Opt.
  Express}\ }\textbf {\bibinfo {volume} {25}},\ \bibinfo {pages} {15938}
  (\bibinfo {year} {2017})}\BibitemShut {NoStop}%
\bibitem [{\citenamefont {Al-Naib}\ \emph {et~al.}(2014)\citenamefont
  {Al-Naib}, \citenamefont {Hebestreit}, \citenamefont {Rockstuhl},
  \citenamefont {Lederer}, \citenamefont {Christodoulides}, \citenamefont
  {Ozaki},\ and\ \citenamefont {Morandotti}}]{PhysRevLett.112.183903}%
  \BibitemOpen
  \bibfield  {author} {\bibinfo {author} {\bibfnamefont {I.}~\bibnamefont
  {Al-Naib}}, \bibinfo {author} {\bibfnamefont {E.}~\bibnamefont {Hebestreit}},
  \bibinfo {author} {\bibfnamefont {C.}~\bibnamefont {Rockstuhl}}, \bibinfo
  {author} {\bibfnamefont {F.}~\bibnamefont {Lederer}}, \bibinfo {author}
  {\bibfnamefont {D.}~\bibnamefont {Christodoulides}}, \bibinfo {author}
  {\bibfnamefont {T.}~\bibnamefont {Ozaki}}, \ and\ \bibinfo {author}
  {\bibfnamefont {R.}~\bibnamefont {Morandotti}},\ }\href {\doibase
  10.1103/PhysRevLett.112.183903} {\bibfield  {journal} {\bibinfo  {journal}
  {Phys. Rev. Lett.}\ }\textbf {\bibinfo {volume} {112}},\ \bibinfo {pages}
  {183903} (\bibinfo {year} {2014})}\BibitemShut {NoStop}%
\bibitem [{\citenamefont {Dai}\ \emph {et~al.}(2004)\citenamefont {Dai},
  \citenamefont {Zhang}, \citenamefont {Zhang},\ and\ \citenamefont
  {Grischkowsky}}]{Dai:04}%
  \BibitemOpen
  \bibfield  {author} {\bibinfo {author} {\bibfnamefont {J.}~\bibnamefont
  {Dai}}, \bibinfo {author} {\bibfnamefont {J.}~\bibnamefont {Zhang}}, \bibinfo
  {author} {\bibfnamefont {W.}~\bibnamefont {Zhang}}, \ and\ \bibinfo {author}
  {\bibfnamefont {D.}~\bibnamefont {Grischkowsky}},\ }\href {\doibase
  10.1364/JOSAB.21.001379} {\bibfield  {journal} {\bibinfo  {journal} {J. Opt.
  Soc. Am. B}\ }\textbf {\bibinfo {volume} {21}},\ \bibinfo {pages} {1379}
  (\bibinfo {year} {2004})}\BibitemShut {NoStop}%
\bibitem [{\citenamefont {Waseda}\ and\ \citenamefont
  {Fujii}(2004)}]{waseda2004density}%
  \BibitemOpen
  \bibfield  {author} {\bibinfo {author} {\bibfnamefont {A.}~\bibnamefont
  {Waseda}}\ and\ \bibinfo {author} {\bibfnamefont {K.}~\bibnamefont {Fujii}},\
  }\href@noop {} {\bibfield  {journal} {\bibinfo  {journal} {Metrologia}\
  }\textbf {\bibinfo {volume} {41}},\ \bibinfo {pages} {S62} (\bibinfo {year}
  {2004})}\BibitemShut {NoStop}%
\bibitem [{\citenamefont {Tsuzuki}\ \emph {et~al.}(2015)\citenamefont
  {Tsuzuki}, \citenamefont {Kuzuu}, \citenamefont {Horikoshi}, \citenamefont
  {Saito}, \citenamefont {Yamamoto},\ and\ \citenamefont {Tani}}]{material}%
  \BibitemOpen
  \bibfield  {author} {\bibinfo {author} {\bibfnamefont {S.}~\bibnamefont
  {Tsuzuki}}, \bibinfo {author} {\bibfnamefont {N.}~\bibnamefont {Kuzuu}},
  \bibinfo {author} {\bibfnamefont {H.}~\bibnamefont {Horikoshi}}, \bibinfo
  {author} {\bibfnamefont {K.}~\bibnamefont {Saito}}, \bibinfo {author}
  {\bibfnamefont {K.}~\bibnamefont {Yamamoto}}, \ and\ \bibinfo {author}
  {\bibfnamefont {M.}~\bibnamefont {Tani}},\ }\href
  {http://stacks.iop.org/1882-0786/8/i=7/a=072402} {\bibfield  {journal}
  {\bibinfo  {journal} {Applied Physics Express}\ }\textbf {\bibinfo {volume}
  {8}},\ \bibinfo {pages} {072402} (\bibinfo {year} {2015})}\BibitemShut
  {NoStop}%
\bibitem [{\citenamefont {Agrawal}(2007)}]{agrawal2007nonlinear}%
  \BibitemOpen
  \bibfield  {author} {\bibinfo {author} {\bibfnamefont {G.~P.}\ \bibnamefont
  {Agrawal}},\ }\href@noop {} {\emph {\bibinfo {title} {Nonlinear fiber
  optics}}}\ (\bibinfo  {publisher} {Academic press},\ \bibinfo {year}
  {2007})\BibitemShut {NoStop}%
\bibitem [{\citenamefont {Deninger}\ \emph {et~al.}(2015)\citenamefont
  {Deninger}, \citenamefont {Roggenbuck}, \citenamefont {Schindler},\ and\
  \citenamefont {Preu}}]{Deninger2015}%
  \BibitemOpen
  \bibfield  {author} {\bibinfo {author} {\bibfnamefont {A.~J.}\ \bibnamefont
  {Deninger}}, \bibinfo {author} {\bibfnamefont {A.}~\bibnamefont
  {Roggenbuck}}, \bibinfo {author} {\bibfnamefont {S.}~\bibnamefont
  {Schindler}}, \ and\ \bibinfo {author} {\bibfnamefont {S.}~\bibnamefont
  {Preu}},\ }\href {\doibase 10.1007/s10762-014-0125-5} {\bibfield  {journal}
  {\bibinfo  {journal} {Journal of Infrared, Millimeter, and Terahertz Waves}\
  }\textbf {\bibinfo {volume} {36}},\ \bibinfo {pages} {269} (\bibinfo {year}
  {2015})}\BibitemShut {NoStop}%
\bibitem [{\citenamefont {Lo}\ and\ \citenamefont {Leonhardt}(2008)}]{lens}%
  \BibitemOpen
  \bibfield  {author} {\bibinfo {author} {\bibfnamefont {Y.~H.}\ \bibnamefont
  {Lo}}\ and\ \bibinfo {author} {\bibfnamefont {R.}~\bibnamefont {Leonhardt}},\
  }\href {\doibase 10.1364/OE.16.015991} {\bibfield  {journal} {\bibinfo
  {journal} {Opt. Express}\ }\textbf {\bibinfo {volume} {16}},\ \bibinfo
  {pages} {15991} (\bibinfo {year} {2008})}\BibitemShut {NoStop}%
\bibitem [{\citenamefont {Vogt}\ and\ \citenamefont
  {Leonhardt}(2017{\natexlab{b}})}]{Vogt:17}%
  \BibitemOpen
  \bibfield  {author} {\bibinfo {author} {\bibfnamefont {D.~W.}\ \bibnamefont
  {Vogt}}\ and\ \bibinfo {author} {\bibfnamefont {R.}~\bibnamefont
  {Leonhardt}},\ }\href {\doibase 10.1364/OE.25.016860} {\bibfield  {journal}
  {\bibinfo  {journal} {Opt. Express}\ }\textbf {\bibinfo {volume} {25}},\
  \bibinfo {pages} {16860} (\bibinfo {year} {2017}{\natexlab{b}})}\BibitemShut
  {NoStop}%
\bibitem [{\citenamefont {Hightower}\ and\ \citenamefont
  {Richardson}(1988)}]{Hightower:88}%
  \BibitemOpen
  \bibfield  {author} {\bibinfo {author} {\bibfnamefont {R.~L.}\ \bibnamefont
  {Hightower}}\ and\ \bibinfo {author} {\bibfnamefont {C.~B.}\ \bibnamefont
  {Richardson}},\ }\href {\doibase 10.1364/AO.27.004850} {\bibfield  {journal}
  {\bibinfo  {journal} {Appl. Opt.}\ }\textbf {\bibinfo {volume} {27}},\
  \bibinfo {pages} {4850} (\bibinfo {year} {1988})}\BibitemShut {NoStop}%
\bibitem [{\citenamefont {Gorodetsky}\ and\ \citenamefont
  {Ilchenko}(1999)}]{Gorodetsky:99}%
  \BibitemOpen
  \bibfield  {author} {\bibinfo {author} {\bibfnamefont {M.~L.}\ \bibnamefont
  {Gorodetsky}}\ and\ \bibinfo {author} {\bibfnamefont {V.~S.}\ \bibnamefont
  {Ilchenko}},\ }\href {\doibase 10.1364/JOSAB.16.000147} {\bibfield  {journal}
  {\bibinfo  {journal} {J. Opt. Soc. Am. B}\ }\textbf {\bibinfo {volume}
  {16}},\ \bibinfo {pages} {147} (\bibinfo {year} {1999})}\BibitemShut
  {NoStop}
\bibitem [{\citenamefont {Naftaly}(2016)}]{naftaly2016international}%
  \BibitemOpen
  \bibfield  {author} {\bibinfo {author} {\bibfnamefont {M.}~\bibnamefont
  {Naftaly}},\ }in\ \href@noop {} {\emph {\bibinfo {booktitle} {Infrared,
  Millimeter, and Terahertz waves (IRMMW-THz), 2016 41st International
  Conference on}}}\ (\bibinfo {organization} {IEEE},\ \bibinfo {year} {2016})\
  pp.\ \bibinfo {pages} {1--2}\BibitemShut {NoStop}%
\bibitem [{\citenamefont {Risti\'{c}}\ \emph {et~al.}(2013)\citenamefont
  {Risti\'{c}}, \citenamefont {Rasoloniaina}, \citenamefont {Chiappini},
  \citenamefont {F\'{e}ron}, \citenamefont {Pelli}, \citenamefont {Conti},
  \citenamefont {Ivanda}, \citenamefont {Righini}, \citenamefont {Cibiel},\
  and\ \citenamefont {Ferrari}}]{Ristic:13}%
  \BibitemOpen
  \bibfield  {author} {\bibinfo {author} {\bibfnamefont {D.}~\bibnamefont
  {Risti\'{c}}}, \bibinfo {author} {\bibfnamefont {A.}~\bibnamefont
  {Rasoloniaina}}, \bibinfo {author} {\bibfnamefont {A.}~\bibnamefont
  {Chiappini}}, \bibinfo {author} {\bibfnamefont {P.}~\bibnamefont
  {F\'{e}ron}}, \bibinfo {author} {\bibfnamefont {S.}~\bibnamefont {Pelli}},
  \bibinfo {author} {\bibfnamefont {G.~N.}\ \bibnamefont {Conti}}, \bibinfo
  {author} {\bibfnamefont {M.}~\bibnamefont {Ivanda}}, \bibinfo {author}
  {\bibfnamefont {G.~C.}\ \bibnamefont {Righini}}, \bibinfo {author}
  {\bibfnamefont {G.}~\bibnamefont {Cibiel}}, \ and\ \bibinfo {author}
  {\bibfnamefont {M.}~\bibnamefont {Ferrari}},\ }\href {\doibase
  10.1364/OE.21.020954} {\bibfield  {journal} {\bibinfo  {journal} {Opt.
  Express}\ }\textbf {\bibinfo {volume} {21}},\ \bibinfo {pages} {20954}
  (\bibinfo {year} {2013})}\BibitemShut {NoStop}%
\end{thebibliography}
\end{document}